\documentclass[conference]{IEEEtran}
\IEEEoverridecommandlockouts
\usepackage{cite}
\usepackage{amsmath,amssymb,amsfonts}
\usepackage{cuted}
\usepackage{algorithmic}
\usepackage{graphicx}
\usepackage{textcomp}
\usepackage{xcolor}
\usepackage{makecell}
\usepackage{verbatim}
\usepackage{cuted}
\usepackage{bm}
\usepackage{stfloats}
\usepackage{amsmath}\allowdisplaybreaks
\def\BibTeX{{\rm B\kern-.05em{\sc i\kern-.025em b}\kern-.08em
    T\kern-.1667em\lower.7ex\hbox{E}\kern-.125emX}}
\usepackage[letterpaper, top=0.75in, bottom=1.1in, left=0.65in, right=0.65in]{geometry}

\begin{document}

\title{Rate-Distortion Region for Distributed Indirect Source Coding with Decoder Side Information\\

}

\author{\IEEEauthorblockN{Jiancheng Tang, Qianqian Yang}

\IEEEauthorblockA{College of information Science and Electronic Engineering, Zhejiang University, Hangzhou 310007, China\\
Email: \{jianchengtang, qianqianyang20\}@zju.edu.cn
}
\thanks{{\thefootnote}{*}This work is partly supported by NSFC under grant No. 62293481, No. 62201505, partly by the SUTD-ZJU IDEA Grant (SUTD-ZJU (VP) 202102).}
}

\maketitle
\begin{abstract}
This paper studies a variant of the rate-distortion problem motivated by task-oriented semantic communication and distributed learning systems, where $M$ correlated sources are independently encoded for a central decoder. The decoder has access to correlated side information in addition to the messages received from the encoders and aims to recover a latent random variable under a given distortion constraint, rather than recovering the sources themselves. We characterize the exact rate-distortion function for the case where the sources are conditionally independent given the side information. Furthermore, we develop a distributed Blahut-Arimoto (BA) algorithm to numerically compute the rate-distortion function. Numerical examples are provided to demonstrate the effectiveness of the proposed approach in calculating the rate-distortion region. THIS PAPER IS ELIGIBLE FOR THE STUDENT PAPER AWARD.

\end{abstract}

\section{Introduction}

\begin{figure}
\centerline{\includegraphics[width=3.0in]{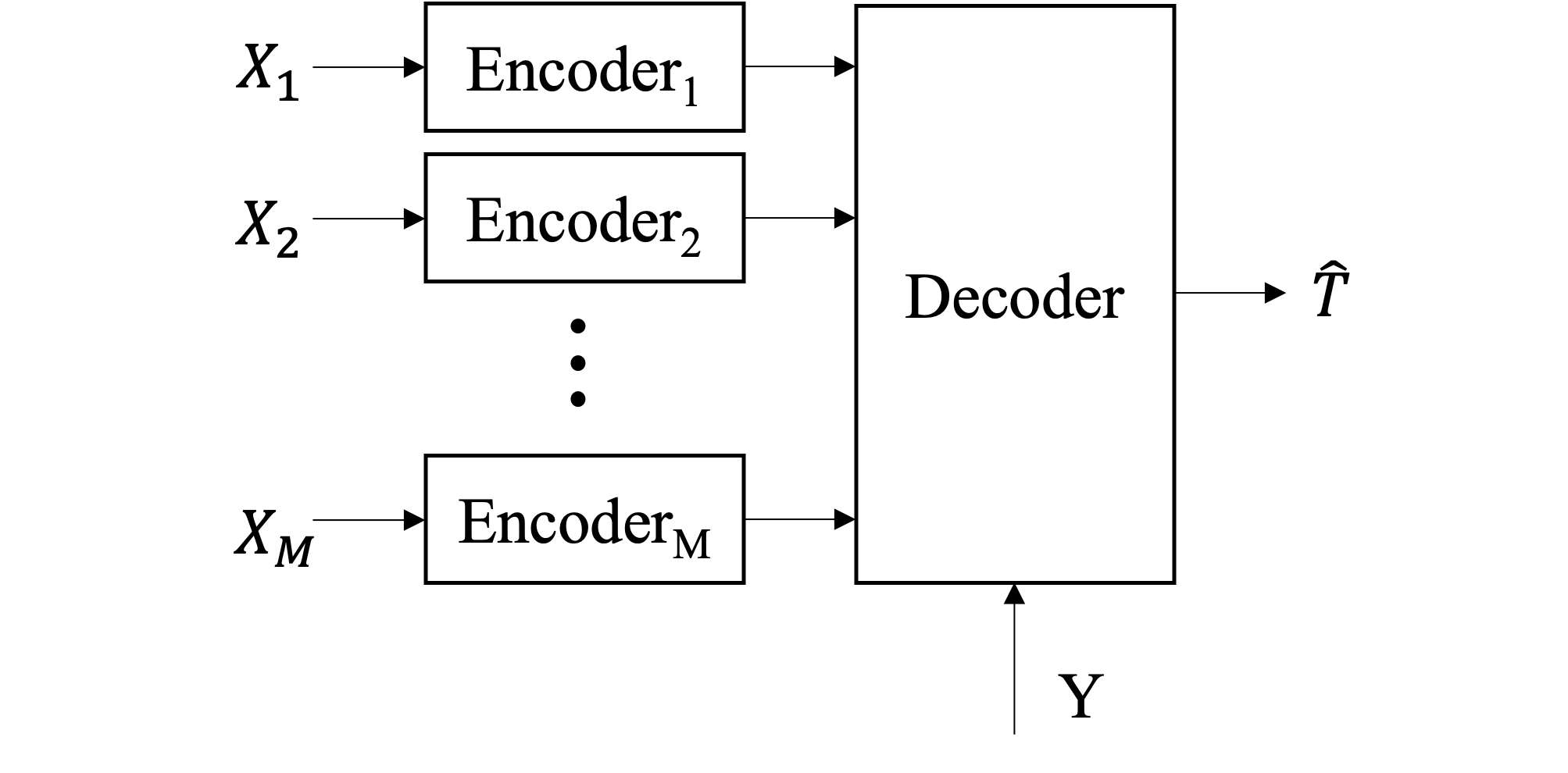}}
\caption{Distributed remote compression of a latent variable with $M$ correlated sources at distributed transmitters and side information at the receiver.}
\label{figsystem}
\end{figure}

Consider the multiterminal source coding setup as shown in Fig.~\ref{figsystem}. Let $(T, X_1,...,X_M,Y)\sim p(t,x_1,...,x_M,y)$ be a discrete memoryless source (DMS) taking values in the finite alphabets $\mathcal{T} \times \mathcal{X}_1 \times \cdots  \times \mathcal{X}_M \times \mathcal{Y}$ according to a fixed and known probability distribution $p(t,x_1,...,x_M,y)$. In this setup, the encoder $m, m\in \mathcal{M}:=\{1,...,M\}$ has local observations 
${X}_m^n:=(X^1_m, \ldots, X^n_m)$. The agents independently encode their observations into binary sequences at rates $\{R_1,\ldots,R_M\}$ bits per input symbol, respectively. The decoder with side information $Y^n = (Y_1, \ldots, Y_n)$ aims to recover some task-oriented latent information ${T}^n:=(T_1, \ldots, T_n)$ which is correlated with $(X^n_1, \ldots, X^n_M)$, but it is not observed directly by any of the encoders. We are interested in the lossy reconstruction of $T^n$ with the average distortion measured by $\mathbb{E}\left[\frac{1}{n}\sum_{i=1}^n d(T_i,\hat{T}_i) \right]$, for some prescribed single-letter distortion measure $d(\cdot, \cdot)$.  A formal $(2^{n R_1},...,2^{n R_M}, n)$ rate-distortion code for this setup consists of
\begin{itemize}
    \item $M$ independent encoders, where encoder $m \in \mathcal{M}$ assigns an index $s_m(x_m^n) \in \left\{1, \ldots, 2^{n R_m}\right\}$ to each sequence $x_m^n \in \mathcal{X}_m^n$;
    \item a decoder that produces the estimate  $\hat{t}^n(s_1,...,s_M,y^n) \in \mathcal{T}^n$ to each index tuple $(s_1,...,s_M)$ and side information $y^n \in \mathcal{Y}^n$.
\end{itemize}

A rate tuple $(R_1,...,R_M)$ is said to be achievable with the distortion measure $d(\cdot, \cdot)$ and the distortion value $D$ if there exists a sequence of $(2^{n R_1},...,2^{n R_M}, n)$ codes that satisfy
\begin{equation}
\begin{aligned}
\mathop {\lim \sup }\limits_{n \to \infty } \mathbb{E}\left[\frac{1}{n}\sum_{i=1}^n d(T_i,\hat{T}_i) \right] \le D.
\end{aligned}
\label{distortion constraint}
\end{equation}
The rate-distortion region $R_{{X_1},\ldots,{X_m}|Y}^{*}\left( {{D}{}} \right)$ for this distributed source coding problem is the closure of the set of all achievable rate tuples $(R_1,\ldots,R_M)$ that permit the reconstruction of the latent variable $T^n$ within the average distortion constraint $D$.


The problem as illustrated in Fig.~\ref{figsystem} is motivated by semantic/ task-oriented communication and distributed learning problems. In semantic/task-oriented communication \cite{han2022semantic}, the decoder only needs to reconstruct some task-oriented information implied by the sources. For instance, it might extract hidden features from a scene captured by multiple cameras positioned at various angles. Here, $T_i$ may also be a deterministic function of the source samples $(X_{1,i}, \ldots, X_{M,i})$, which then reduces to the problem of lossy distributed function computation \cite{two, howto, distributed}. A similar problem also arises in distributed training. Consider $Y^n$ as the global model available at the server at an iteration of a federated learning process, and $(X^n_1, \ldots, X^n_M)$ as the independent correlated versions of this model after downlink transmission and local training. The server aims to recover the updated global model, $T^n$, based on the messages received from all $M$ clients. It is often assumed that the global model is transmitted to the clients intact, but in practical scenarios where downlink communication is limited, the clients may receive noisy or compressed versions of the global model \cite{amiri2020federated, gruntkowska2024improving, Amiri:TWC:22}. 

For the case of $M=1$, the considerd problem reduces to the remote compression in a point-to-point scenario with side information available at the decoder. In \cite{Photios1, liu}, the authors studied this problem without the correlated side information at the receiver, motivated in the context of semantic communication. This problem is known in the literature as the remote rate-distortion problem \cite{Dobrushin, Wolf:TIT:70}, and the rate-distortion trade-off is fully characterized in the general case. The authors studied this trade-off in detail for specific source distributions in \cite{Photios1}. Similarly, the authors of \cite{Guo} characterized the remote rate-distortion trade-off when correlated side information is available both at the encoder and decoder. Our problem for $M=1$ can be solved by combining the remote rate-distortion problem with the classical Wyner-Ziv rate-distortion function \cite{wyner1, wyner2}.

The rate-distortion region for the multi-terminal version of the remote rate-distortion problem considered here remains open. Wagner \emph{et al.} developed an  improved outer bound for a general multi-terminal source model \cite{wagner}. Sung \emph{et al.} proposed an achievable rate region for the distributed lossy computation problem, without giving an conclusive rate-distortion function\cite{Sung}. Gwanmo \emph{et al.} considered a special case in which the sources are independent and derived a single-letter expression for the rate-distortion region \cite{Gwanmo}. Gastpar \cite{Michael} considered the lossy compression of the source sequences in the presence of side information at the receiver. He characterized the rate-distortion region for the special case, in which $X_i$'s are conditionally independent given the side information.


To provide a performance reference for practical coding schemes, it is necessary not only to characterize the exact theoretical expression for the rate-distortion region but also to calculate the rate-distortion region for a given distribution and a specific distortion metric. In the traditional single-source direct scenario, determining the rate-distortion function involves solving a convex optimization problem, which can be addressed using the globally convergent iterative Blahut–Arimoto algorithm, as discussed in \cite{BAcompute}. In this paper, we are interested in computing the rate-distortion region $R_{{X_1},\ldots,{X_m}|Y}^{*}\left( {{D}{}} \right)$ for the general distributed coding problem. We pay particular attention to the special case in which the sources are conditionally independent given the side information, motivated by the aforementioned examples. 

In Section II, we give a single letter information theoretic expression for the rate-distortion region $R_{{X_1},\ldots,{X_m}|Y}^{*}\left( {{D}{}} \right)$. In Section III, we develop an alternating minimization framework to calculate the rate-distortion region by generalizing the Blahut–Arimoto (BA) algorithm. In Section IV, we demonstrate the effectiveness of the proposed framework through numerical examples. 

\section{A Single Letter Expression for the Rate Region}

In this section, we first introduce an achievable rate region $R_a\left( {D} \right) $, which is contained within the goal rate-distortion region $R_a\left( {{D}} \right) \subseteq R_{{X_1},{...},{X_M}|Y}^{*}\left( {{D}} \right)$. Then, we derive a region $R_o\left( {{D}} \right) $ which contains the  rate-distortion region $R_o\left( {{D}} \right)  \supseteq  R_{{X_1},{...},{X_M}|Y}^{*}\left( {{D}} \right)$. We show that the two regions coincide and characterize the rate-distortion function exactly
under the scenario where the sources are conditionally independent given the side information.

\emph{Lemma 1:} $R_a\left( {{D}} \right) \subseteq R_{{X_1},{...},{X_M}|Y}^{*}\left( {{D}} \right)$, where $R_a\left( {{D}} \right)$ is the set of all rate tuples $(R_1,...,R_M)$ that satisfy the following conditions
\begin{equation}
\begin{aligned}
\sum_{i \in A} R_i \geq I(\bm{X}_A; \bm{W}_A \mid \bm{W}_{A^c}, Y) \quad \\\text{for all } A \subseteq \{1, \ldots, M\},
\end{aligned}
\label{innerbound}
\end{equation}
and there exist a decoder ${g}\left(  \cdot  \right)$  such that
\begin{equation}
\begin{aligned}
E{d}({T},{g(\bm{W},Y)}) \leqslant {D},\\
\end{aligned}
\label{constrains}
\end{equation}
where $A^c$ includes all elements in the set $ A \subseteq \{1, \ldots, M\}$ that are not in  $A$,

\emph{Proof:} This can be proven by extending the achievable region proof for the Wyner-Ziv problem with multiple sources, as established by Gastpar (Theorem 2, \cite{Michael}). However, we omit the detailed proof here due to space limitations.


\emph{Lemma 2:} $R_o\left( {{D}} \right)  \supseteq  R_{{X_1},{...},{X_M}|Y}^{*}\left( {{D}} \right)$, where $R_o\left( {{D}} \right) $ is the set of all rate triples $(R_1,...,R_M)$ for which the following conditions are satisfied
\begin{equation}
\begin{aligned}
 \sum_{i \in A} R_i \geq \sum_{i \in A} I(X_i; W_i \mid  Y) \quad \\ \text{for all } A \subseteq \{1, \ldots, M\},
\end{aligned}
\label{outerbound}
\end{equation}
and there exist a decoding function ${g}\left(  \cdot  \right)$ such that

\begin{equation}
\begin{aligned}
E{d}({T},{g_1(\bm{W},Y)}) \leqslant {D}.\\
\end{aligned}
\label{constrains}
\end{equation}

\emph{Theorem 3:} If  $X_1, ..., X_M$ are conditionally independent given the side information $Y$, the achievable region $R_a\left( {D} \right) $ coincides with  $R_o\left( {{D}} \right) $, i.e.,
\begin{equation}
\begin{aligned}
R_a\left( {{D}} \right) = R_o\left( {{D}} \right)  =  R_{{X_1},{...},{X_M}|Y}^{*}\left( {{D}} \right).
\end{aligned}
\label{constrains}
\end{equation}

\emph{Proof:} Lemma 2 and Theorem 3 can be easily established based on Wagner's proof, see \cite{wagner}.

\emph{Corollary 4:} If $X_1, ..., X_M$ are conditionally independent given the side information $Y$,  the expression of the rate-distortion region \eqref{innerbound} and \eqref{outerbound} can be simplified to
\begin{equation}
\begin{aligned}
 R_i \geq  I(X_i; W_i) -I( W_i ;  Y) \quad \text{for all } l \in \{1, \ldots, M\}.
\end{aligned}
\label{coincide}
\end{equation}

\emph{Proof:} The proof of Corollary 4 easily follows from the conditional independence property. The rigorous proof will be provided in a longer version.

\section{Iterative optimization framework based on BA algorithm}
In this section, we present the iterative optimization framework for calculating the rate distortion region. Starting with the standard Lagrange multiplier method, the problem of calculating the rate-distortion region $R_{{X_1},\ldots,{X_m}|Y}^{*}\left( {{D}{}} \right)$ \eqref{coincide} is equivalent to minimize
\begin{equation}
 \sum_{i \in \mathcal{M} := \{1, \ldots, M\}} I(W_i; X_i) - I(W_i; Y) + \lambda (E[d(T, \hat{T})]-D)
 \label{start1}
\end{equation}
By the definition of mutual information, we can rewrite \eqref{start1} as
\begin{equation}
\begin{aligned}
&\mathcal{L}_{\lambda }(\bm{Q}, \bm{q}, {q}') \\&= \sum_{y, \hat{t}, x_i, w_i,i \in \mathcal{M}} 
p( y, x_i) q_i(w_i \mid x_i) q'(\hat{t} \mid y,\bm{w}) 
\log \frac{q_i(w_i \mid x_i)}{Q_i(w_i \mid y)}
\quad \\&
+ \lambda \sum_{\bm{w}, \bm{x}, t, \hat{t},y} d(t, \hat{t}) p(t,\bm{x},y) q'(\hat{t} \mid y,\bm{w}) 
\prod_{i \in \mathcal{M}} q_i(w_i \mid x_i),
\end{aligned}
\label{1}
\end{equation}
where $\bm{Q}, \bm{q}, {q}'$ represent the distributions that need to be iteratively updated, and the vectorized notation $\bm{Q}$ represents the conditional distribution of the auxiliary variables given $Y$, i.e., $[Q_i(w_i \mid y) \mid w_i \in \mathcal{W}_i, y \in \mathcal{Y}, i \in \mathcal{M}]$,  $\bm{q}$ represents the conditional distribution of the auxiliary variables given sources $X_i$, $[q_i(w_i \mid x_i) \mid  w_i \in  \mathcal{W}_i, x_i \in \mathcal{X}_i, i \in \mathcal{M}]$ and ${q}'$ represents the conditional distribution of the indirect variable $T$ given $Y$ and auxiliary variables, $q'(\hat{t} \mid y, w_1,...,w_M)$. 

\emph{Lemma 5 (Optimization of $\bm{Q}$)}: For a fixed  $\bm{Q}_{\backslash m}, \bm{q}, q'$, the Lagrangian $\mathcal{L}_{\lambda }(\bm{Q}, \bm{q}, {q}')$ is minimized by
\begin{equation}
Q_{m}^*(w_m \mid y) \triangleq 
\frac{
\sum_{ x_m, \hat{t}} p( y, x_m) q(w_m \mid  x_m) q'(\hat{t} \mid y, \bm{w})
}{
\sum_{ x_m, \hat{t}, w_m} p( y, x_m) q(w_m \mid  x_m) q'(\hat{t} \mid y, \bm{w}),
}    
\end{equation}
where $\bm{Q}_{\backslash m} = \left[ Q_i(w_i \mid y) \mid  w_i \in \mathcal{W}_i, y \in \mathcal{Y}, i \in \mathcal{M} \backslash m \right]$.

\emph{Proof:}  For any $Q_m$
\begin{equation}
\begin{small}
\begin{aligned}
&\mathcal{L}_{\lambda }({Q}_{m}^*, \bm{q}, {q}') -\mathcal{L}_{\lambda }({Q}_{m}, \bm{q}, {q}') 
\\&= \sum_{y, \hat{t}, x_m, w_m} 
p( y, x_m) q_i(w_m \mid x_m) q'(\hat{t} \mid y,\bm{w}) 
\log \frac{q_m(w_m \mid x_m)}{Q_{m}^*(w_m \mid y)}
\\& -\sum_{y, \hat{t}, x_m, w_m} 
p( y, x_m) q_i(w_m \mid x_m) q'(\hat{t} \mid y,\bm{w}) 
\log \frac{q_m(w_m \mid x_m)}{Q_{m}(w_m \mid y)}
\\& =\sum_{y, \hat{t}, x_m, w_m} 
p( y, x_m) q_i(w_m \mid x_m) q'(\hat{t} \mid y,\bm{w}) 
\log \frac{Q_{m}(w_m \mid y)}{Q_{m}^*(w_m \mid y)}
\\& \overset{(a)}{\leq} \sum_{y, \hat{t}, x_m, w_m} 
p( y, x_m) q_i(w_m \mid x_m) q'(\hat{t} \mid y,\bm{w}) 
 (\frac{Q_{m}(w_m \mid y)}{Q_{m}^*(w_m \mid y)}-1)
 \\&=0,
\end{aligned}
\end{small}
\label{Q}
\end{equation}
where  (a) follows from the fact that $\log (1+x) \leq x$, and the equality is achieved if $Q_{m}^* = Q_{m}$. This completes the proof of Lemma 5.

\begin{figure*}[hb]
	\centering
	\vspace*{8pt}
	\hrulefill
	\vspace*{20pt} 
	\begin{eqnarray}
q^*(w_m \mid  x_m) = 
\frac{
\exp \left[ \sum_{y} p(y \mid x_m) \log Q(w_m \mid y) 
- \lambda \sum_{\bm{w}_{\backslash m}, \bm{x}_{\backslash m}, t, \hat{t},y} d(t, \hat{t}) p(t,\bm{x},y) q'(\hat{t} \mid y,\bm{w}) \bm{q}_{\backslash m}(\bm{w}_{\backslash m} \mid \bm{x}_{\backslash m}) \right]
}{
\sum_{w_m} \exp \left[ \sum_y p(y \mid x_m) \log Q(w_m \mid y) 
- \lambda \sum_{\bm{w}_{\backslash m}, \bm{x}_{\backslash m}, t, \hat{t},y} d(t, \hat{t}) p(t,\bm{x},y) q'(\hat{t} \mid y,\bm{w}) \bm{q}_{\backslash m}(\bm{w}_{\backslash m} \mid \bm{x}_{\backslash m})\right]
}
    \label{q}
    \end{eqnarray}
\end{figure*}

\emph{Lemma 6 (Optimization of $\bm{q}$)}: For a fixed  $\bm{Q}, \bm{q}_{\backslash m}, q'$, the Lagrangian $\mathcal{L}_{\lambda }(\bm{Q}, \bm{q}, {q}')$ is minimized by \eqref{q} shown at the bottom of the page, and the minimum is given by
\begin{equation}
\small
\begin{aligned}
&\mathcal{L}_{\lambda }(\bm{Q}, \bm{q}_{\backslash m}^*, {q}') \\&=
\sum_{x_i,i \in \mathcal{M}} p(x_i) 
\min_{w_i}
[
\sum_{y, \hat{t}} p(y \mid  x_i) q'(\hat{t} \mid y, \bm{w}) 
\log \frac{q^*(w_i \mid  x_i)}{Q(w_i \mid y)}
\\&+ \lambda \sum_{\bm{w}, \bm{x}, t, \hat{t},y} d(t, \hat{t}) p(t,\bm{x},y) q'(\hat{t} \mid y,\bm{w}) 
\prod_{i \in \mathcal{M}} q_i(w_i \mid x_i)
].
\end{aligned}
\end{equation}

\emph{Proof:}  For a fixed  $\bm{Q}, \bm{q}_{\backslash m}, q'$, the Lagrangian $\mathcal{L}_{\lambda }(\bm{Q}, \bm{q}, {q}')$ is minimized by $q^*(w_m \mid  x_m)$ if and only if the following Kuhn-Tucker(KT) conditions are satisfied
\begin{equation}
\left. \frac{\partial \mathcal{L}_{\lambda }}{\partial q_m} \right|_{q_m^*} = \gamma, \quad \text{if } q^*(w_m \mid x_m) > 0 ,
\label{kt1}
\end{equation}
and
\begin{equation}
\left. \frac{\partial \mathcal{L}_{\lambda }}{\partial q_m} \right|_{q_m^*} \leq \gamma, \quad \text{if } q_m^*(w_m \mid x_m) = 0 .
\end{equation}

Since 
\begin{equation}
\begin{aligned}
&\frac{\partial \mathcal{L}_{\lambda }}{\partial q_m} \\&= 
\sum_{ x_m, y} p(x_m,y) \left(\log \frac{p(w_m \mid x_m)}{p(w_m \mid y)} +1\right) \\&+ \lambda\sum_{\bm{w}, \bm{x}, t, \hat{t},y} d(t, \hat{t}) p(t,\bm{x},y) q'(\hat{t} \mid y,\bm{w}) 
q_{\backslash m}(\bm{u}_{\backslash m} \mid \bm{x}_{\backslash m}),
\end{aligned}
\end{equation}
the first KT condition \eqref{kt1} becomes
\begin{equation}
\small
\begin{aligned}
&\tilde{\gamma}\\& = \sum_{ x_m, y} p(x_m,y) \left(\log {p(w_m \mid x_m)} - \log {p(w_m \mid y)}\right) \\&+ \lambda\sum_{\bm{w}_{\backslash m}, \bm{x}_{\backslash m}, t, \hat{t},y} d(t, \hat{t}) p(t,\bm{x},y) q'(\hat{t} \mid y,\bm{w}) 
q_{\backslash m}(\bm{w}_{\backslash m} \mid \bm{x}_{\backslash m}),
\end{aligned}
\end{equation}
where 
\begin{equation}
    q_{\backslash m}(\bm{w}_{\backslash m} \mid \bm{x}_{\backslash m}) = \prod_{i \in \mathcal{M}\backslash m} q_i(w_i \mid x_i)
\end{equation}
and we have
\begin{equation}
\begin{aligned}
&q(w_m \mid  x_m) \\&=  \exp{\left(\frac{ \tilde{\gamma}}{p(x_m)} \right)} \exp ( \sum_{y} p(y \mid x_m) \log Q(w_m \mid y) 
\\&- \lambda \sum_{\bm{w}_{\backslash m}, \bm{x}_{\backslash m}, t, \hat{t},y} d(t, \hat{t}) p(t,\bm{x},y) q'(\hat{t} \mid y,\bm{w}) {q}_{\backslash m}(\bm{w}_{\backslash m} \mid \bm{x}_{\backslash m})  ).
\end{aligned}
\end{equation}
Then, \eqref{q} can be obtained after normalizing $q(w_m \mid  x_m) $.

\emph{Lemma 7 (Optimization of $q'$)}: For a fixed  $\bm{Q}, \bm{q}$, the Lagrangian $\mathcal{L}_{\lambda }(\bm{Q}, \bm{q}, {q}')$ is minimized by the maximum a Bayes detector.
\begin{equation}
q'^*(\hat{t} \mid \bm{w}, y) =
\begin{cases} 
\alpha(\bm{w}), & \hat{t} \in \arg\min_{\hat{t} \in \hat{\mathcal{T}}} \mathbb{E}[d(t, \hat{t}) \mid \bm{W} = \bm{w}], \\
0, & \text{otherwise}
\end{cases}
\end{equation}
where $\alpha(\bm{w})$ is selected to guarantee
\begin{equation}
\sum_{\hat{t}}q'^*(\hat{t} \mid \bm{w}, y)=1
\end{equation}
and $\mathbb{E}[d(t, \hat{t}) \mid \bm{W} = \bm{w}]$ denotes
\begin{equation}
\sum_{\bm{w}_{\backslash m}, \bm{x}_{\backslash m}, t, \hat{t},y} d(t, \hat{t}) p(t,\bm{x},y) 
\prod_{i \in \mathcal{M}} q_i(w_i \mid x_i)
\end{equation}

\emph{Proof:} We note that the distortion term in the Lagrangian $\mathcal{L}_{\lambda }$ that depends on $q'(\hat{t} \mid y,\bm{w}) $ is the mean distortion, which can be minimized by a Bayes detector.

Based on Lemmas 5, 6 ,7, we have the iterative algorithm for computing the rate-distortion region for the distributed indirect source with decoder side information. For a  given $i \in \mathcal{M}$, we iteratively calculate the following \eqref{iterative Q} and \eqref{iterative q} to alternately update $Q_i$ and $q_i$
\begin{equation}
Q_i^{\ell,k_i}= 
\frac{
\sum_{ x_i, \hat{t}} p( y, x_i) q^{(\ell,k_i)}(w_i \mid  x_i) q'(\hat{t} \mid y, \bm{w})
}{
\sum_{ x_i, \hat{t}, w_i} p( y, x_i) q^{(\ell,k_i)}(w_i \mid  x_i) q'(\hat{t} \mid y, \bm{w})
}  
\label{iterative Q}
\end{equation}
\begin{figure*}[hb]
	\centering
	\vspace*{8pt}
	\hrulefill
	\vspace*{20pt} 
	\begin{eqnarray}
    \footnotesize
q_i^{(l,k_i+1)} = 
\frac{
\exp \left[ \sum_{y} p(y \mid x_i) \log Q^{(l,k_i)}(w_i \mid y) 
- \lambda \sum_{\bm{w}_{\backslash i}, \bm{x}_{\backslash i}, t, \hat{t},y} d(t, \hat{t}) p(t,\bm{x},y) q'^{(\ell,i)}(\hat{t} \mid y,\bm{w}) \bm{q}_{\backslash i}^{(\ell,i)}(\bm{w}_{\backslash i} \mid \bm{x}_{\backslash i}) \right]
}{
\sum_{w_i} \exp \left[ \sum_y p(y \mid x_i) \log Q^{(\ell,k_i)}(w_i \mid y) 
- \lambda \sum_{\bm{w}_{\backslash i}, \bm{x}_{\backslash i}, t, \hat{t},y} d(t, \hat{t}) p(t,\bm{x},y) q'^{(\ell,i)}(\hat{t} \mid y,\bm{w}) \bm{q}_{\backslash i}^{(\ell,i)}(\bm{w}_{\backslash i} \mid \bm{x}_{\backslash i})\right]
}
    \label{iterative q}
    \end{eqnarray}
\end{figure*}
until the Lagrangian $\mathcal{L}_{\lambda }$ converges, and the associated $q_i$ is
\begin{equation}
q_i^{(\ell, *)}(w_i \mid x_i) = \lim_{k_i \to \infty} q_i^{(\ell, k_i)}(w_i \mid x_i) =: q_i^{(\ell+1, 1)}.
\end{equation}
Then we update $q'(\hat{t} \mid \bm{w}, y)$ according to
\begin{equation}
q'^{\ell',i'} =
\begin{cases} 
\alpha^{\ell',i'}(\bm{w}), & \hat{t} \in \arg\min_{\hat{t} \in \hat{\mathcal{T}}}  \bm{J}^{(\ell,i)}( \hat{t} , \bm{w}), \\
0, & \text{otherwise}
\end{cases}
\end{equation}
with $i' = i + 1$ and $\ell'=\ell$ if $i < M$ and $i'=1$ and $\ell'=\ell+1$ if $i=M$, and where $\alpha^{\ell',i'}(\bm{w})$ is selected to guarantee
\begin{equation}
\sum_{\hat{t}}q'^{(\ell',i')}(\hat{t} \mid \bm{w}, y)=1
\end{equation}
and 
\begin{equation}
\begin{aligned}
\bm{J}^{(\ell,i)}( \hat{t} , \bm{w}) &= \sum_{ \bm{x}, t, y} d(t, \hat{t}) p(t,\bm{x},y) \\& \times q_{\backslash i}^{(\ell,i)}(w_{\backslash i} \mid x_{\backslash i}) q_i^{(\ell,*)}(w_i \mid x_i)
\end{aligned}
\end{equation}

Next we repeat the process for the next user, i.e., $(i \gets i + 1 \text{ if } i < M \text{ and } i \gets 1, \ell \gets \ell + 1 \text{ if } i = M)$.

\emph{Convergence analysis:}

The algorithm employs an alternating minimization approach that produces Lagrangian values that are monotonically non-increasing and bounded below, thereby generating a convergent sequence of Lagrangians. Since the rate component of the Lagrangian is convex, and the sum of convex functions remains convex, the Lagrangian will also be convex if the expected distortion is a convex function. As a result, the proposed iterative optimization framework is capable of achieving the global minimum \cite{elements}.

\begin{equation}
\sum_{\bm{w}, \bm{x}, t, \hat{t},y} d(t, \hat{t}) p(t,\bm{x},y) q'(\hat{t} \mid y,\bm{w}) \prod_{i \in \mathcal{M}} q_i(w_i \mid x_i)
\label{distortion}
\end{equation}

We note that the expected distortion \eqref{distortion} includes a product of variables in the optimization process, it is not a linear function, and the Lagrangian may exhibit non-convex behavior.  Even when the problem is non-convex, the authors in \cite{computing} demonstrate that  the BA-based iterative algorithm initialized randomly and followed by selecting the minimum Lagrangian among all converged values can still provide highly effective information-theoretic inner bounds for the rate-distortion region, serving as a benchmark for practical quantization schemes.

\section{NUMERICAL EXAMPLES}

In this section, we provide an example to illustrate the proposed iterative algorithms for computing the rate-distortion region of a distributed source coding problem with decoder side information. As in the problem considered in this paper, distributed edge devices compress their observations $\{X_1,...,X_M\}$ and transmit them to a central server (CEO). The central server then aims to recover the indirect information $T$ from the received data, utilizing side information $Y$. For the convenience of demonstration, we consider a simple case where $M = 2$ and the sources are binary, i.e., $\mathcal{X}_1=\mathcal{X}_2=\mathcal{Y}=\{0, 1\}$. The joint distributions, denoted by $Q(x_1, y)$ and $Q(x_2, y)$ are given by
\begin{equation}
\begin{aligned}
Q(x_1, y) = \frac{(1 - p_1)}{2}\delta_{x_1,y} + \frac{p_1}{2}(1 - \delta_{x_1,y}), 
\\Q(x_2, y) = \frac{(1 - p_2)}{2}\delta_{x_2,y} + \frac{p_2}{2}(1 - \delta_{x_2,y}), 
\end{aligned}
\end{equation}
where the Kronecker delta function $\delta_{x, y}$ equals 1 when x = y, and 0 otherwise. We can consider $Y$ as the input to two different binary symmetric channels (BSCs) with crossover probabilities $p_1, p_2$, respectively, where $0 \leq p_i \leq \frac{1}{2}, i \in \{1,2\}$. The corresponding outputs of these channels are $X_1$ and $X_2$. In this example, we set $p_1 = p_2 = 0.3$.

We also assume that the information of interest is directly the combination of the two distributed sources, i.e., $T = \{X_1, X_2\}$. The distortion measure is given by $d(t, \hat{t})=d(x_1, \hat{x}_1)+d(x_2, \hat{x}_2)$, where
\begin{equation}
d(x_i, \hat{x}_i) =
\begin{cases} 
0, & x_i = \hat{x}_i, \\
1, & x_i \neq \hat{x}_i.
\end{cases}
\end{equation}

\begin{figure}
\centerline{\includegraphics[width=3.5in]{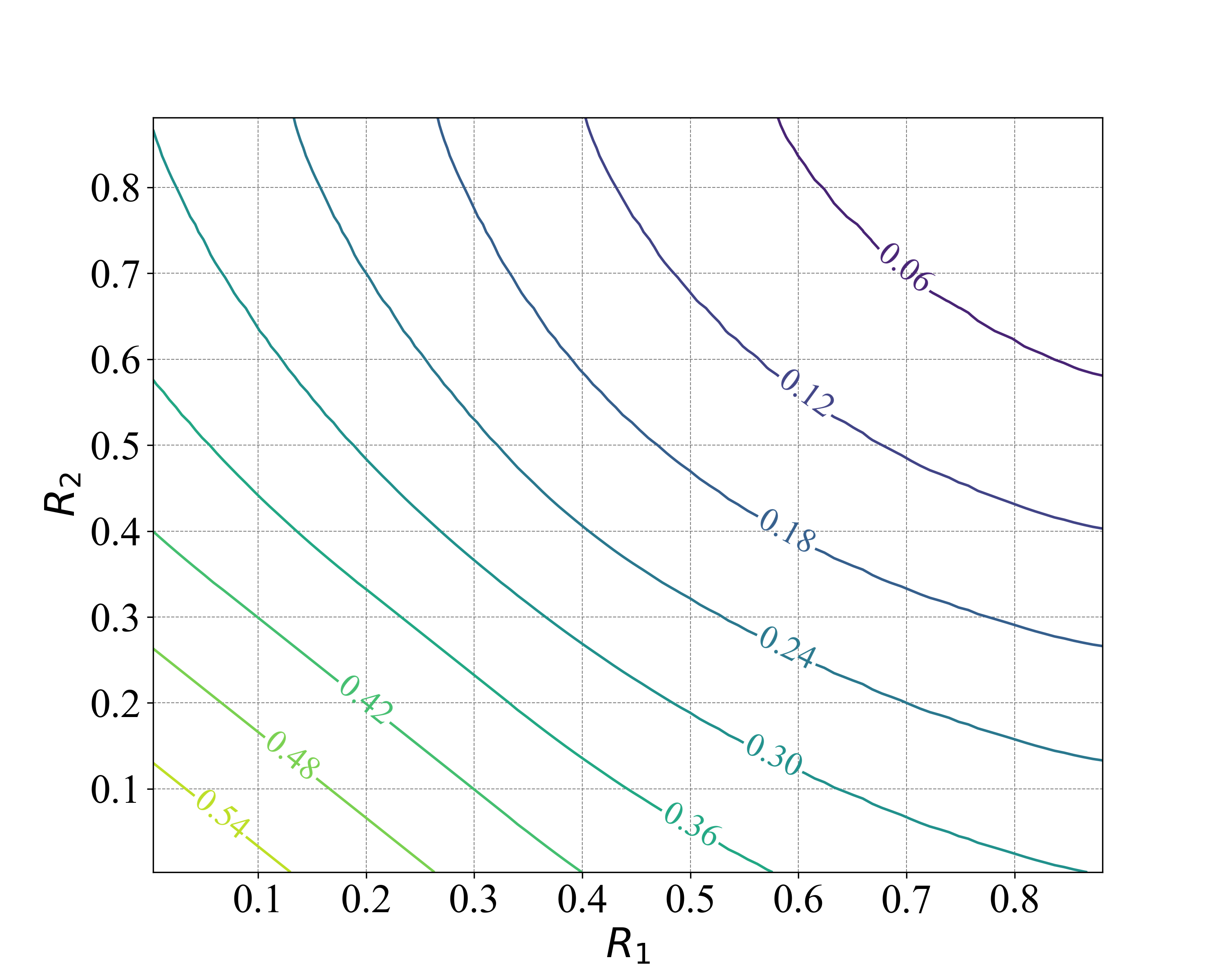}}
\caption{Contour plot of the rate distortion region with two distributed binary sources $\{X_1, X_2\}$, where the labels on the contours represent the distortion values $D$ on $d(t, \hat{t})$.}
\label{contour}
\end{figure}

\begin{figure}
\centerline{\includegraphics[width=3.5in,trim=3cm 1cm 3cm 3cm, clip]{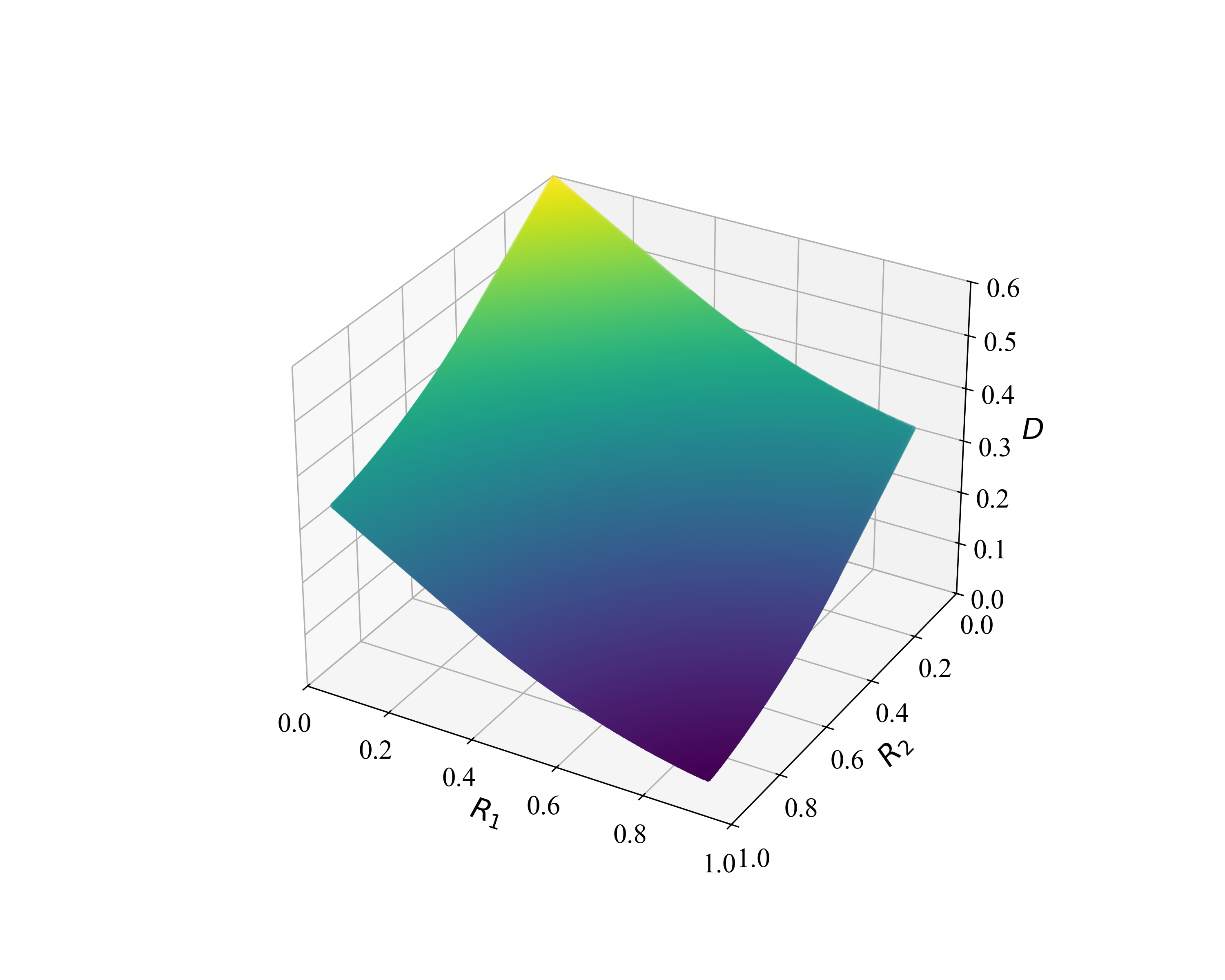}}
\caption{Surface plot of the rate-distortion region.}
\label{Surface}
\end{figure}

By applying the proposed iterative optimization framework, we can obtain the optimal transition probability distribution $Q_i^*(w_i \mid y),q_i^*(w_i \mid x_i)$ and $q'^*(\hat{t} \mid \bm{w}, y)$ that meets a given distortion constraint $D$ on $d(t, \hat{t})$, and the corresponding  minimum rate $R_i$ can be calculated by
\begin{equation}
\begin{aligned}
   R_i = \sum_{y, x_i, w_i} 
p( y, x_i) q_i^*(w_i \mid x_i)  
\log \frac{q_i^*(w_i \mid x_i)}{Q_i^*(w_i \mid y)}
\quad
\end{aligned}
\end{equation}

\begin{figure}
\centerline{\includegraphics[width=3.7in]{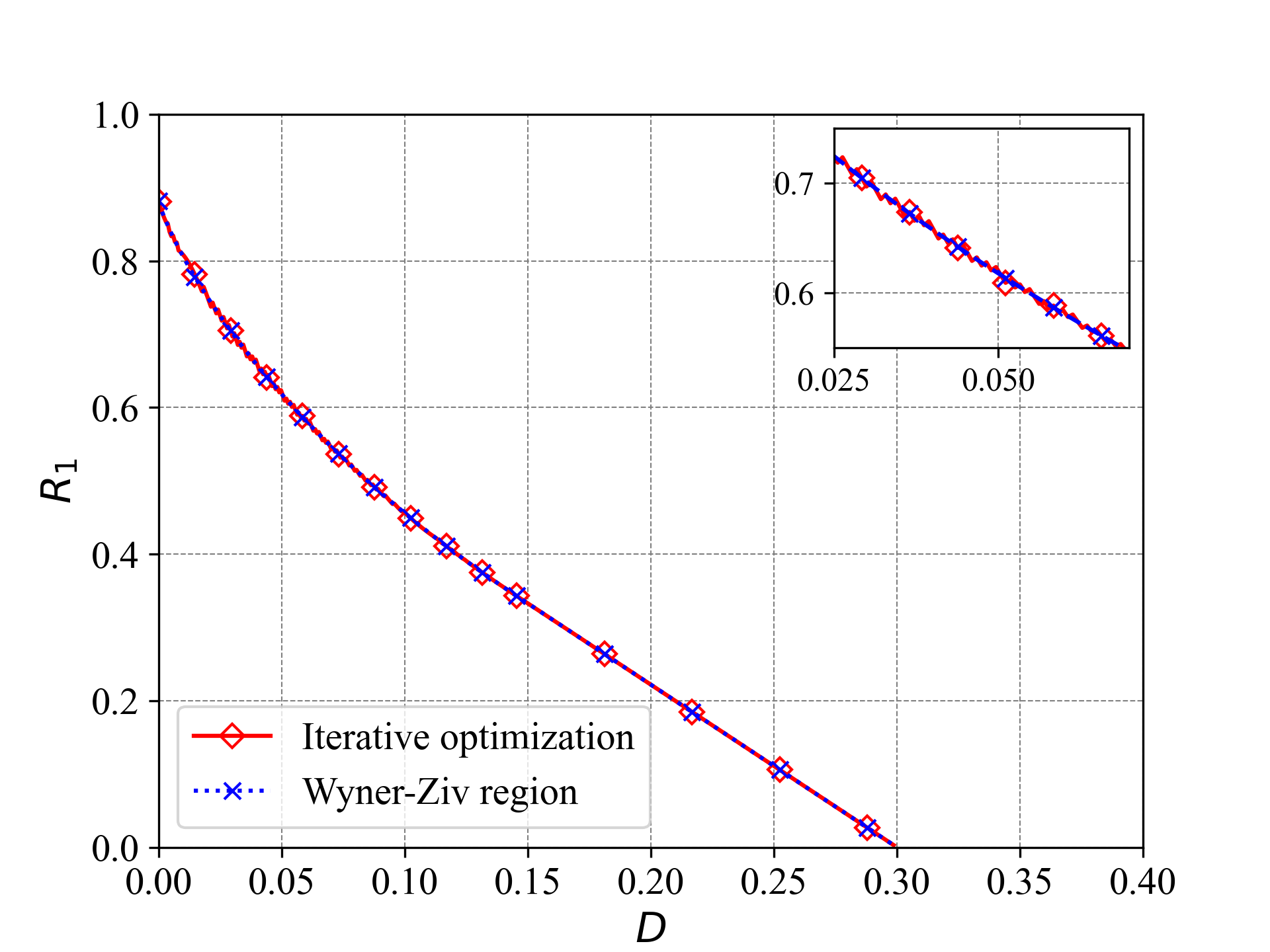}}
\caption{The rate-distortion function for the case when $M=1$, i.e., the Wyner-Ziv problem}
\label{single}
\end{figure}

The contour plot of the rate-distortion region for this scenario is presented in Fig.~\ref{contour}, while Fig.~\ref{Surface} displays a surface plot depicting the rate-distortion region. We note that when $M=1$, the considered problem reduces to the traditional point-to-point Wyner-Ziv problem. In Fig.~\ref{single}, we compare the rate-distortion results computed using the proposed approach with the theoretical analysis by Wyner \emph{et al.} \cite{wyner1976rate}. We observe that the two rate-distortion function curves coincide, demonstrating the effectiveness of the proposed iterative approach for calculating rate distortion.


\section{Conclusion}
This paper explored a variant of the rate-distortion problem motivated by semantic communication and distributed learning systems, where correlated sources are independently encoded for a central decoder to reconstruct the indirect source of interest. In addition to receiving messages from the encoders, the decoder has access to correlated side information and aims to reconstruct the indirect source under a specified distortion constraint. We derived the exact rate-distortion function for the case where the sources are conditionally independent given the side information. Furthermore, we introduced a distributed iterative optimization framework based on the Blahut-Arimoto (BA) algorithm to numerically compute the rate-distortion function. A numerical example has been provided to demonstrate the effectiveness of the proposed approach.

\clearpage
\bibliographystyle{IEEEtran}
\bibliography{ref}

\end{document}